# Development of nanostructured-graphene-supported silver nanoparticles as catalysts for electroreduction of oxygen in alkaline electrolyte

*Sylwia Zoladek[a], Magdalena Blicharska[a], Agnieszka Anna Krata[a], Iwona A. Rutkowska[a], Anna Wadas[a], Krzysztof Miecznikowski[a], Enrico Negro[b], Vito Di Noto[b], Pawel J. Kulesza[a]\**

[a] Department of Chemistry, University of Warsaw, Pasteura 1, PL-02-093 Warsaw, Poland

[b] Department of Industrial Engineering, Università degli Studi di Padova in Department of Chemical Sciences, Via Marzolo 1, 35131 Padova (PD), Italy

\* corresponding authors
E-mail: pkulesza@chem.uw.edu.pl, tel: +48228220211 ext. 218 or 308, fax: +48228225996

**Keywords:**

Polyoxometallates; $SiO_2$-modified Graphene, Chemically reduced graphene oxide, Graphene, Silver nanoparticles; Electrocatalysis; Oxygen reduction; Alkaline medium

**Abstract**

Here we develop a class off face centred cubic structure of metallic silver nanocrystals, which are enriched with high-index facets, to enable high ORR activity process. Silver nanoparticles deposited within different carbon supports: carboxylated-graphene, $SiO_2$-doted reduced-graphene-oxide ($Gr/SiO_2$) and reduced-graphene-oxide (Gr) were prepared by the chemical reduction method, in which the Keggin-type phosphotugstate heteropolyblue was used as reducing agent for the precursor ($AgNO_3$). The reduction of silver ions and the formation of silver nanoparticles deposited within graphene-type supports have been assessed by UV-Vis spectroscopy and was confirmed by XRD studies, which revealed that the AgNPs were crystalline in nature. The major advantage of the proposed chemical synthetic method is the integration of the superb properties of both silver nanoparticles and graphene supports in a single-step synthesis with a 100% usage of the silver precursor ($AgNO_3$) according to the coupled plasma mass spectrometer (ICP-MS). The choice of the carbon support strongly affected catalytic activity of the resulting Ag nanoparticles towards oxygen electroreduction in alkaline medium. We show that $SiO_2$-doted reduced- graphene-oxide supported silver nanoparticles display significantly enhanced catalytic activity towards the oxygen reduction reaction (ORR) in alkaline solutions compared to the silver nanoparticles immobilized within carboxylated-graphene support and chemically reduced graphene oxide (Gr). Functionalization of graphene with $SiO_2$ provides an efficient way to keep the nanosheets of graphene exfoliated, thus making them more easily accessible for the intercalation. Moreover under RRDE conditions the hybrid material based on $SiO_2$-doted reduced- graphene-oxide supported silver nanoparticles displays the most impressive electrocatalytic performance toward ORR and shows the highest number of exchanged electrons (n) ranging from 3.96 to 3.998 when compared to the silver nanoparticles immobilized within carboxylated-graphene support (3.90-3.994) and reduced-graphene-oxide (Gr) supported silver nanoparticles (3.88-3.991).

# 1. Introduction

Recently, alkaline fuel cells (AFCs) have regained attention because, compared to the acid environment of a proton exchange membrane fuel cell (PEMFC), systems mounting hydroxyl ($OH^-$)- conducting polymer membranes provide reduced alcohol crossover [1] and a less corrosive environment to the catalysts and electrodes [2]. Intense research efforts have been focused on the development of non-platinum catalyst to replace Pt for catalyzing oxygen reduction reaction (ORR), and to reduce the cost of the fuel cell production [3-10].

The AFCs open up the possibility of using wider choice of the catalyst, including: Co and Fe phthalocyanines [3], manganese dioxide [11], $Co_3O_4$/nitrogen-doped graphene hybrids [12] as well as less expensive metal catalysts including silver [4] and gold [5] what makes the alkaline fuel cell a potentially low cost technology compared to acid direct alcohol fuel cell technology, which employs platinum catalysts. The inherently faster kinetics of the oxygen reduction reaction in an alkaline fuel cell allows the use of non-noble and low-cost metal electrocatalysts such as silver nanoparticles showing a promising potential to be an alternative to platinum candidates for implementation as the oxygen reduction catalytic sites for oxygen reduction reaction (ORR) under alkaline conditions [13,14].

Blizanac et al. postulated that the kinetic of oxygen electroreduction is determined by the availability of free active sites on Ag(hkl) surfaces in alkaline media and that the ORR is a structure-sensitive reaction [15]. They reported that the structure-sensitivity arises from different strengths of adsorption of OH and $O_2$ on various Ag facets order of activities of Ag single-crystal planes towards the ORR.

According to Wang [16] the oxygen reduction proceeds one-step "direct" four-electron reduction on silver nanodecahedra, while two-step processes on silver nanocubes. Moreover the weaker adsorption of OH* on silver (111) facet provides more active sites, leading to the higher catalytic activity of ORR on silver nanodecahedra than that on silver nanocubes [16]. The simulations results suggest that the different ORR catalytic activity can be interpreted by the adsorption competition between oxygen and hydroxyl on different silver facets.

According to Lu [17] the unusual catalytic activity of silver nanoparticles may be accounted for by the high fraction of surface atoms and their low coordination numbers, which may be manipulated readily by the nanoparticle dimensions. Moreover there has been carried out research to evaluate the effect of the Ag particles size on their catalytic activity towards oxygen electroreduction. It was found that the Ag NPs sized less than 3nm exhibited high electrocatalytic activity for oxygen reduction [18]. The later phenomenon was attributed to the narrow gap between the d-band and the Fermi level of low-coordinated metal atoms, allowing for easier adsorption of oxygen molecules on the cluster surface when compared the close-packed counterparts. Also the type of capping agent adsorbed on silver seems to be crucial factor. Compton and co-workers have demonstrated that citrate-capped Ag NPs showed higher peroxide production than bulk Ag [19].

Designing catalysts with both optimal activity and stability for ORR in alkaline solutions remain outstanding challenges. Under such conditions, when metallic nanoparticles are used alone (without protecting layers and appropriate support) the parasite effects related to agglomeration and degradation of catalytic nanoparticles are likely to be largely prevented. Moreover the capping agents on the surface of the clusters may block the mass transport and electron transfer, which seriously impair their electroactivity. What is more the extremely small size and high surface energy, the durability of metal clusters is largely restricted by the easy occurrence of dissolution, aggregation, and sintering during catalysis reactions [20]. To overcome these obstacles, a plausible solution is to anchor metallic nanoparticles on the specific supports, which not only prevent catalyst against agglomeration and corrosion but also interact with catalytic particles deposited on the surface, allowing to control and manipulate the electrocatalytic behaviour making it favourable for achieving improved oxygen reduction kinetics [21]. Regarding to the excellent properties of graphene-based materials, including long-term stability, facilitating dispersion of metallic nanoparticles, providing easy access of reactants, and assure good electrical contact with active sites graphene-type materials seem to be optimal supports for metallic nanoparticles [22].

In the present study, we consider three kinds of catalytic systems, namely silver nanoparticles deposited within different carbon supports: carboxylated-graphene, $SiO_2$-doted reduced-graphene-oxide (Gr/$SiO_2$) and reduced-graphene-oxide (Gr) for oxygen electroreduction in alcaline medium. All of catalytic systems were prepared by the chemical reduction method, in which the particularly reduced Keggin-type phosphotugstates were used as reducing and stabilizing agent for the controlled reduction of precursor (AgNO$_3$). Catalyst structure, morphology, and electrochemical properties were characterized by X-ray diffraction (XRD), UV-Vis absorption spectroscopy, transmission electron microscopy (TEM), and cyclic voltammetry (CV) techniques. Cyclic voltammetry and rotating ring disk electrode methods were employed to investigate the catalytic activity of the resulting catalysts for the oxygen electro-reduction. Influences of different type of graphene supports on the on the ORR mechanisms were investigated. It was found that the oxygen reduction on Gr/$SiO_2$ supported silver nanoparticles precedes the most efficient four-electron reduction. The high electrocatalytic activity was assigned to the presence of high fraction of Ag (111) facets. Moreover a thorough kinetics analysis performed for the AgNPs/Gr/$SiO_2$ catalysts confirmed that all prepared materials containing Ag can be successfully used for the preparation of cathode catalyst for the ORR.

The incorporation of the corrosion-resistant Gr/$SiO_2$, as the support for Ag nanoparticles is a possible strategy to achieve long-term operation of FCs. Due to its remarkable thermal, mechanical and electrical properties and ease of functionalization, $SiO_2$ -modified graphene materials are able to prevent corrosion and activity loses of Ag catalyst. Moreover functionalization of graphene with $SiO_2$ provides an efficient way to keep the nanosheets of graphene exfoliated, thus making them more easily accessible for the interactions with precursor ions and $SiW_{12}O_{40}^{4-}$ capping agents.

## 2. Experimental

Chemicals were commercial materials of the highest available grade, and they were used as received. KOH, ethanol, methanol and $K_3[Fe(CN)_6]$ were obtained from POCh (Poland). The Silver nitrate, $AgNO_3$ (99.9999%); sodium borohydride (powder, 98%), $NaBH_4$; silicotungstic acid hydrate, $H_4[Si(W_3O_{10})_4] \cdot xH_2O$, 5 wt% Nafion solution were purchased from Sigma-Aldrich and were used without any further purification. Nitrogen and oxygen gases (purity 99.999%) were from Air Products (Poland). Reduced graphene oxide (rGO) was obtained using sodium borohydride as reducing agent according to the procedure described earlier [23].

The preparation of the $SiO_2$-doted reduced-graphene-oxide (rGO-$SiO_2$) nanocomposite was carried out through a proprietary procedure [24, 25].

Chemicals and materials used for determination of total content of Ag in graphene samples by ICP-MS: Trace analysis grade reagents of 96% sulphuric acid (Suprapur, Merck, Germany) and 65% nitric acid (Suprapur, Merck, Germany) were used for digestion of sample. Silver single component standard ICP (TraceCERT, Merck, Germany) with concentration of 10 mg $L^{-1}$ during the measurements by calibration curve ICP-MS was used.

The working standard solutions of silver were prepared gravimetrically using an analytical balance (Mettler Toledo, Switzerland). High quality deionised water from Milli-Q system (Millipore, Merck, Germany) was used throughout this work.

The high-resolution thermogravimetric profiles were collected between 30 and 1000°C by means of a TGA 2950 analyzer (TA instruments). The latter has a resolution of 1 g; the sensitivity ranges between 0.1 and 2% $min^{-1}$, on the basis on the first derivative of the weight loss. The heating ramp falls between 50 and 0.001°C $min^{-1}$, depending on the first derivative of the weight loss. The profiles were measured by using samples of *ca.* 5 mg, that were placed on an open Pt pan. Measurements were collected in an oxidizing atmosphere of dry air. Wide-angle X-ray diffraction (WAXD) profiles were measured with a diffractometer manufactured by GNR, mod. eXplorer; the instrument mounted a monochromatized CuK source. The angles were calibrated using a standard NIST Si 640d; the instrumental line broadening was accounted for using a reference $Y_2O_3$ sample, that was annealed in air. WAXD patters were acquired between 3 and 70°, using a 0.05° step and an integration time of 30 sec.

The spectral Raman profiles were measured with a confocal Raman microscope (model DRX, Thermo Scientific), which uses an excitation laser having a wavelength of 532 nm. Transmission electron microscopy (TEM) pictures, both conventional and at high resolution, were obtained with a JEOL 3010 microscope equipped with a high-res pole piece yielding a point-by-point resolution of 0.17 nm. The instrument also mounted a Gatan slowscan 794 CCD camera. The sample preparation protocol for TEM was described in the literature [26].

Inductively coupled plasma mass spectrometer NexION 300D (Perkin Elmer, USA) was used for the ICP-MS measurements. The optimized working conditions used for silver determination are listed in Table 1.

**Table 1.** ICP-MS optimized working parameters.

| Parameter | Description |
| --- | --- |
| Spray chamber | Cyclonic, Scott type |
| Nebulizer | Quartz coaxial, type Meinhard |
| Plasma torch | Quartz |
| Sampler cone | Nickel |
| Skimmer cone | Nickel |
| Hyper skimmer cone | Aluminium |
| Plasma gas flow | 16 L min$^{-1}$ |
| Nebulizing gas flow | 0.8 L min$^{-1}$ |
| RF power | 1400 W |
| Analog detector voltage | -1800 V |
| Pulse detector voltage | 950 V |
| Lens voltage | -9 V |
| Dead time of detector | 25 ns |
| Sample flow rate | 1 mL min$^{-1}$ |
| Numer of sweeps | 5 |
| Number of repeats | 6 |
| Integration time | 150 ms |
| Monitored isotopes | $^{107}$Ag, $^{109}$Ag |

A microwave closed digestion unit (Ultrawave, Milestone, USA) was used for the digestion. Digestion was performed for around 20 mg of material with addition of 1.5 mL 96% $H_2SO_4$ and 0.5 mL 65% $HNO_3$. After digestion samples were diluted till 30 mL with deionized water. Digestion program was as followed:

**Table 2.** Digestion program parameters:

| Step | Time [min] | Power [W] | T1 [°C] | T2 [°C] | Pressure [bar] |
|---|---|---|---|---|---|
| 1 | 00:30:00 | 1500 | 250 | 60 | 140 |
| 2 | 00:30:00 | 1500 | 250 | 60 | 140 |

The procedural blank sample was prepared in the same way as sample with Ag. For the determination of total content of Ag in graphene samples and procedural blank sample, the inductively coupled plasma mass spectrometer (ICP-MS) and external calibration curve with 3 different concentrations of Ag standard solutions were used.

Electrochemical measurements were carried out on CH Instruments (Austin, TX, USA) Models: 600B and 750A workstations. The electrochemical cell was assembled with a conventional three-electrode system: a rotating ring disk working electrode (RRDE), saturated calomel electrode (SCE) reference electrode (exhibiting potential of ca. 240 mV relative to the reversible hydrogen electrode (RHE)) and carbon wire counter electrode. The rotating ring disk electrode (RRDE) working assembly was from Pine Instruments; it included a glassy carbon (GC) disk and a Pt ring. The radius of the GC disk electrode was 2.5 mm; and the inner and outer radii of the platinum ring electrode were 3.25 and 3.75 mm, respectively. Before experiments, working electrode was polished with aqueous alumina slurries (grain size, 5–0.05 mm) on a Buehler polishing cloth. Later, the glassy carbon disk electrode was subjected to potential cycling in 0.1 mol $dm^{-3}$ KOH for 30 min in the potential range from 0 to 1.0 V. The Pt ring electrode was also subjected to potential cycling in the same solution but in the range of the potential form -0.1 V to 1.2 V. The collection efficiency (N) of the RRDE assembly, was calibrated by $K_3Fe(CN)_6$ redox reaction determined according to procedure described before [27]. During the RRDE experiments in oxygen saturated solutions, the potential of the ring electrode was kept at 1.2 V vs. RHE. The activity of the prepared catalytic films was evaluated by performing voltammetric potential cycles in the range from 0.1 to 1.1 V vs. RHE at 50 mVs$^{-1}$ until stable voltammetric responses were observed. In order to obtain the representative background responses typical cyclic voltammograms were recorded by scanning potential from 0.1 to 1.1 V in 0.1 mol dm$^{-3}$ KOH at a scan rate of 10 mVs$^{-1}$ under nitrogen atmosphere. Before the electrochemical measurement the electrolyte was purged and saturated with $N_2$ or $O_2$ gas. A constant nitrogen (or oxygen) flow over the solution was maintained during all measurements. All RRDE polarization curves were recorded at the scan rate of 10 mV s$^{-1}$, typically with a rotation rate of 1600 rpm. The ring potential was maintained at 1.21 V vs RHE to oxidize any hydrogen peroxide produced. Experiments were performed at room temperature (22 ± 2°C).

Fabrication of silver nanoparticles modified with heteropolytungstates ($SiW_{12}O_{40}^{-4}$) supported onto: carboxylated-graphene or $SiO_2$-doted reduced- graphene-oxide ($Gr/SiO_2$) or reduced-graphene-oxide (Gr) was performed according to the manner described below. First 0.3 g of carboxylated-graphene or $SiO_2$-doted reduced- graphene-oxide ($Gr/SiO_2$) or reduced-graphene-oxide (Gr) was suspended in the 460 ml of ethanol and, subsequently, subjected to 24 h stirring to obtain homogenous mixture. Then 0.270 g of $H_4[Si(W_3O_{10})_4] \cdot xH_2O$ and 0.2 g of $AgNO_3$ were added to the each suspension. The resulting solution had been magnetically stirred for 24 h; and, later, each suspension was cooled down to the temperature of 2 °C and was purged and saturated with $N_2$ gas. A constant nitrogen flow over the solution was maintained during the synthesis. To obtain a silver loading on the level of 30 wt% of Ag on the appropriate amound of freshly prepared aqueous solution of sodium tetrahydroborate ($NaBH_4$) was added (prepared by addition of 0.5 g $NaBH_4$ into 1 cm$^3$ of distilled water). The solutions were mixed for 24 h at temperature of 2 °C to ensure the reaction completeness. Later each suspension was filtered, washed out three times with water and dried in a vacuum oven at 80 °C for 12 h, and then the catalyst sample was obtained.

The morphology of samples was characterized by Libra transmission electron 120 EFTEM (Carl Zeiss) operating at 120 kV.

Ultraviolet-visible absorption analysis (250-800 nm) of the following solutions (suspensions): solution of $H_4SiW_{12}O_{40}$ with $SiO_2$-doted reduced- graphene-oxide, solution of partially reduced $H_4SiW_{12}O_{40}$ (heteropolyblue), $SiO_2$-doted reduced-graphene-oxide, silver nanoparticles supported onto $H_4SiW_{12}O_{40}$-modified $SiO_2$-doted reduced- graphene-oxide and reduced-graphene-oxide (rGO) supported silver nanoparticles subsequently cleaned from their inorganic capping agents ($PW_{12}O_{40}^{3-}$) by converting to 0.1 moldm$^{-3}$ water solution of KOH were recorded with UV-Vis model Lambda Spectrophotometer of Perkin Elmer.

X-Ray diffraction (XRD) spectrum was collected using Rigaku Ultima IV diffractometer using Co-Kα radiation ($\lambda = 0.179$ nm).

A catalytic inks containing silver nanoparticles supported onto carboxylated-graphene, $SiO_2$-doted reduced graphene-oxide (AgNPs/Gr/$SiO_2$) and reduced-graphene-oxide (AgNPs/Gr) were prepared according to the procedure as follows: 6,57 mg of silver nanoparticles supported onto carboxylated-graphene or AgNPs/Gr/$SiO_2$ or AgNPs/Gr was suspended in the 460 ml of ethanol and, subsequently, subjected to 24 h stirring to obtain homogenous mixture. Then 40 ml of Nafion (5% alcoholic solution) was added to the suspension. The resulting solution had been magnetically stirred for 24 h; As a rule, appropriate amounts of the resulting catalytic inks were dropped onto surfaces of glassy carbon electrodes to obtain loadings of silver nanoparticles equal to 30 mg cm$^{-2}$ and, later, subjected to drying in air at room temperature for 30 min. For comparison, inks containing only bare carboxylated-graphene, $SiO_2$-doted reduced- graphene-oxide (Gr/$SiO_2$) and reduced-graphene-oxide (Gr) were prepared according to analogous procedures, except in the absence of silver nanoparticles. In other words, 4.6 mg of carboxylated-graphene or $SiO_2$-doted reduced or graphene-oxide (Gr/$SiO_2$) or reduced-graphene-oxide (Gr) was suspended in the 460 ml of ethanol and 40 ml of Nafion (5% alcoholic solution). Later 2 ml of the appropriate bare carbon suspension was introduced onto the surface of the glassy carbon disk electrode; and the suspension was air-dried at room temperature.

## 3. Results and discussion

3.1. *Physicochemical Identity of SiO$_2$-modified Reduced Graphene Oxide and Reduced Graphene Oxide*

In this study, we fabricated silver nanoparticles supported on three kinds of graphene-type materials: carboxylated-graphene, reduced-graphene-oxide (Gr) and SiO$_2$-doted reduced-graphene-oxide (Gr/SiO$_2$) showing a potential to adsorb and store water molecules due to the carbon-oxygen groups (hydroxyl, epoxy, carbonyl, carboxyl) on their hydrophilic surface. Independent analysis based on the high-resolution thermogravimetric analysis (HR-TGA) profiles (Fig. 1a) showed that the mass loss at T < 100°C, that is ascribed to the removal of atmospheric moisture, is equal to *ca.* 0.1 and 2 wt% for reduced-graphene-oxide (Gr) and Gr/SiO$_2$, respectively. These results indicate that both samples are not very hygroscopic; in detail, Gr/SiO$_2$ is slightly more hygroscopic than Gr, possibly owing to the presence of polar, hydrophilic SiO$_2$ structures. Moreover both Gr/SiO$_2$ and pristine Gr exhibit only one main degradation event, that is associated to the oxidative decomposition of graphene sheets. This event takes place at *ca.* 690 and 667°C for Gr and Gr/SiO$_2$, respectively. In this latter sample, the temperature of the main decomposition event is slightly lower, owing to the presence of: (a) a high concentration of defects on the Gr sheets (*cfr.* (Fig. 1c) and (Fig. 2c)); and (b) SiO$_2$, that is expected to facilitate the adsorption of oxygen in the system, triggering the oxidative decomposition event.

The high-temperature residue of Gr and Gr/SiO$_2$ is equal to *ca.* 0.4 and 73 wt%. Correspondingly the amount of carbon included in the samples, that undergoes decomposition during the HR-TGA measurements, is equal to: (i) more than 99% (in the case of Gr); and (ii) *ca.* 25 wt% (in the case of Gr/SiO$_2$).

The features of the wide-angle X-ray diffraction (WAXD) pattern of Gr (Fig. 1b) match quite closely those of graphitic C (S.G. *P*6$_3$/mc, COD#9008569) [28]; in particular, a strong peak is revealed at 2θ ~ 26.6°, ascribed to the residual stacking of graphene sheets along the [001] direction. The WAXD pattern of Gr/SiO$_2$ reveals a broad halo centered at 2θ ~ 24°, assigned to the amorphous silica nanoparticles (NP) included in the system. No sharp peak was revealed at 2θ ~ 26.6°, witnessing that in Gr/SiO$_2$ the graphene sheets are almost completely exfoliated.

The SiO$_2$-doted reduced-graphene-oxide (Gr/SiO$_2$) material has also been characterized (relative to reduced-graphene-oxide (Gr)) by Raman spectroscopy (Fig. 1c). The Raman spectra of the samples (Fig.1c) reveal features that are easily assigned on the basis of the literature [29-31]. In particular, the Raman spectra show two large peaks: one located near 1350 cm$^{-1}$ which is attributed to the D band originating from the amorphous structures of carbon, and the second one close 1580 cm$^{-1}$, which is correlated with the G band and reflects the graphitic structures of carbon. I$_D$/I$_G$ ratio of Gr and Gr/SiO$_2$ is equal to 0.06 and 1.34, respectively. This evidence indicates that the surface concentration of interfacial defects on the Gr sheets included in Gr/SiO$_2$ is significantly higher than that of pristine Gr. The lower degree of Gr/SiO$_2$ surface organization when compared to Gr is very promising for further applications in electrocatalysis.

Figure 2 clearly indicates that Gr/SiO$_2$ consists of highly defective Gr sheets covering/wrapping the SiO$_2$ NPs (appearing as dark features in Figure 2(b)). The high-resolution micrographs further support this evidence; in detail: (i) Figure 2(a) reveals directly the highly defective edges of the Gr nanoplatelets included in Gr/SiO$_2$; and (ii) Figure 2(b) witnesses that Gr nanoplatelets cover/wrap the SiO$_2$ NPs.

## 3.2. Formation and morphology of Graphene-supported silver nanoparticles

Three types of catalytic systems based on graphene-supported silver nanoparticles nanoparticles were prepared by the chemical reduction method following three stages, in which the $NaBH_4$-prereduced Keggin-type silicotungstate heteropolyblue was acting as the reducing agent for the precursor ($AgNO_3$). The spectrophotometry in visible region has been applied to monitor simultaneously every stage of formation of silicotungstate-protected silver nanoparticles immobilized onto $SiO_2$-doted reduced- graphene-oxide (Gr/$SiO_2$) (Fig. 3). Addition of sodium tetrahydroborate to the solution of $H_4SiW_{12}O_{40}$ and Gr/$SiO_2$ results in a rapid change of color from brown (Fig. 3a, dask line) to dark blue (Fig. 3b, dotted line) indicating that heteropoly blue was generated, which led to visible characteristic broad tail extending from 500 to well-beyond 800 nm with two absorption peaks located at 490 and 740 nm, attributed to the d-d bands at 490 nm and intervalence charge transfer (IVCT, W(V) → W(VI)) bands at about 625-800 nm [32]. Injection of $AgNO_3$ precursor solution to the partially-reduced mixed-valence W(VI,V) heteropolyblue solution caused depleted of spectrum assigned as the inter-valence charge-transfer (Fig. 3c, solid line). Moreover the appearance of a new prominent and sharp absorption band was observed to form at about 400 nm (Fig. 3c) and should be correlated with the characteristic of silver nanoparticles arising due to excitation of surface plasmon vibrations in the Ag nanoparticles with spherical structure [33, 34].

The dispersion and morphology of the $SiO_2$-doted reduced- graphene-oxide (Gr-$SiO_2$) supported silver nanoparticles, carboxylated-graphene-immobilized Ag nanoparticles and reduced-graphene-oxide (Gr) supported Ag nanoparticle was identified by the Transmission Electron Microscopy (TEM). From Fig. 4A we see a uniform coating of spherical silver nanoparticles on the Gr-$SiO_2$ sheet. An inset in Fig. 4A' shows the histogram of the diameter of nanoparticles indicates the average diameter of the silver nanoparticles to be around 5 nm. Morphology studies of carboxylated-graphene-immobilized Ag nanoparticles (Figure 4B) predominantly show the formation of higher population of small nanoparticles sized between 1-6 nm, moreover TEM have revealed the presence less uniform silver nanoparticles sized from 7 to 28 nm dispersed within within the representative graphene sheet (Figure 4B, 4B'). The presence of a large number of smaller silver nanoparticles adhering to Gr is probably a result of the second nucleation processed of precursor agent. The explanation of these phenomena should take into account the presence of sufficient amount of different type of surface sites including: unsaturated carbons defect sites, surface carbon-oxygen polar groups and partly reduced interfacial layers of $SiW_{12}O_{40}^{4-}$ sites moderate reducing capability of carbon support providing excessive nucleation sites during nucleation of precursor. It is apparent from Fig. 4C and 4C that Ag nanoparticles deposited onto reduced-graphene-oxide are approximately 4 nm in diameter, uniformly dispersed within Gr and du to the presence of attractive precursor–interface interaction no agglomerations were obtained.

The total content of silver on different graphene supports was determined by the calibration curve ICP-MS. For the determination of total content of Ag deposited on different graphene-type supports and procedural blank sample, the inductively coupled plasma mass spectrometer (ICP-MS) and external calibration curve with 3 different concentrations of Ag standard solutions were used. According to ICP-MS Ag content in different Graphene-based sample was on the level of 28% ± 4% (2*SD). This evidence indicates that the proposed chemical synthetic method allowed for 100% usage of the silver precursor ($AgNO_3$).

XRD was used to examine the crystalline phase of silver nanostructures deposited on of three different supports: carboxylated-graphene, $SiO_2$-doted reduced- graphene-oxide (Gr/$SiO_2$) and reduced-graphene-oxide (Gr) and are shown in Figure 5. The peaks present in all three XRD profiles located at about 38.3°, 44.2°, 64.4° and 77.4° should be attributed to the (111), (200), (220), and (311) crystalline planes of Ag with a face-centered cubic (fcc) structure [35-37].

To comment the electrochemical properties of AgNPs/Gr/$SiO_2$ relative to carboxylated-graphene supported Ag NPs and AgNPs/Gr the cyclic voltammograms were collected in 0.1 moldm$^{-3}$ KOH solution saturated by N$_2$ bubbling. Dotted lines stand for responses of the respective Ag-free graphene-based supports. Briefly, all catalytic systems: AgNPs/Gr/SiO$_2$ (as for Fig. 6A), carboxylated-graphene supported Ag NPs (as for Fig. 6B) and AgNPs/Gr (as for Fig. 6C) show quite high specific capacitance attributed to both high specific surface area of the catalytic systems (electric double layer capacitance) and redox processes of carbons (pseudocapacitance). The cyclic voltammetric behavior of SiO$_2$-doted reduced- graphene-oxide (Fig. 6A□) and the carboxylated-graphene (Fig. 6B□) show higher impact of pseudocapacitance exhibited by the large and nonrectangular CV curve when compared to Gr (Fig. 6C□) attributed to the presence of higher fraction polar groups. An important observation inferred from Figure 5 is that all three cyclic voltammograms standing for responses of the catalytic layers containing Ag display a very high increase of background currents originating from the double-layer-type charging/discharging effects at potential over 0.6 V, which confirm that incorporation of silver nanoparticles into catalytic layers changes the electrochemical properties of graphene. All three working electrodes modified with AgNPs/Gr/SiO$_2$ (as for Fig. 6A), carboxylated-graphene supported Ag NPs (as for Fig. 6B) and AgNPs/Gr (as for Fig. 6C) showed strong Ag redox pair at potential over 0.8 V anodic peaks attributed to electrochemical adsorption of OH and formation of oxide layer and cathodic peak attributed to the electrochemical reduction of Ag$_2$O and Ag(OH) formed under anodic conditions into metallic Ag [15,16,38]. It is interesting to see that the reduction potential of silver oxides on AgNPs/Gr/SiO$_2$ is shifted towards more positive potentials when compared to carboxylated-graphene supported Ag NPs and AgNPs/Gr which confirms that Ag is in the metallic state during oxygen electroreduction process.

To support our view that silicotungstates protecting layers (used for the synthesis of silver nanoparticles) are completely removed from the graphene-supported silver nanoparticles after convering Gr supported Ag nanoparticles (initially modified with H$_4$SiW$_{12}$O$_{40}$) into alkaline solution of 0.1 moldm$^{-3}$ KOH we have subjected ultraviolet-visible absorption analysis (Fig. 7). It is apparent from the Curve (Fig. 7a) that before such treatment the UV-vis spectrum of the silver nanoparticles supported onto $H_4SiW_{12}O_{40}$-modified reduced- graphene-oxide is characterized by one peak with maximum absorption wavelength value of 310 nm, which can be assigned to the oxygen–tungsten charge transfer transition [39,40]. It is noteworthy that absorption band with maximum at 310 nm band disappeared after in solution after alkali treatment, which proves an efficient decomposition of polyoxometalate capping monolayer-type films (Fig. 7b). This finding could be explained by the fact that the Keggin anions were completely decomposed into $SiO_4^{4-}$ and $WO_4^{2-}$ in alkaline solution of pH = 13. It is clearly observed that the absorption maximum at about 400 nm attributed to the excitation of surface plasmon vibrations in the Ag nanoparticles remains unchanged and no broadening of the absorption band is observed, indicating no aggregation of the silver nanoparticles consequent to removal of the silicotungstates protecting layers [41].

3.3. *Electroreduction of oxygen*

To investigate the electrocatalytic activities of as prepared catalysts, the diagnostic linear scan voltammetric experiments in $O_2$-saturated 0.1 moldm$^{-3}$ KOH solutions at different modified compounds: AgNPs/Gr/SiO$_2$, carboxylated-graphene supported AgNPs and AgNPs/Gr were measured at a constant active mass loading and all the electrodes showed a substantial oxygen reduction process in the presence of oxygen (Fig. 8). Comparison of results in Fig. 8 leads to the conclusion that, electrode modified with AgNPs/Gr/SiO$_2$ exhibit greatly positive shift in the ORR and the peak appearing at the most positive potential (0.81 V in Fig. 8A) when compared to carboxylated-graphene supported AgNPs with the ORR peak potentials at 0.77 V (Fig. 8B, Fig. 8C). The highest current densities for oxygen electroreduction process are as follows: 0.58 mAcm$^{-2}$ (AgNPs/Gr/SiO$_2$ in Fig. 8A), 0.57 mAcm$^{-2}$ (carboxylated-graphene supported AgNPs in Fig. 8B) and 0.56 mAcm$^{-2}$ (AgNPs/Gr in Fig. 8C) indicating that the choice of carbon support strongly effect the electrocatalytic performance. It is undoubted that the choice of SiO$_2$-doted reduced- graphene-oxide for silver nanoparticles deposition plays a key role for improved ORR activity. This behavior occurred from the presence of SiO$_2$ improving the water retention, thus leading to enhanced conductivity and better oxygen electroreduction performance [42].

3.4. *Diagnostic RRDE experiments*

Further the hydrodynamic voltammetry (RRDE) experiments were carried out to evaluate the mechanism of ORR for: Gr/SiO$_2$ supported silver nanoparticles (as for Fig. 9A), carboxylated-graphene supported silver nanoparticles (as for Fig. 9B) and reduced graphene oxide supported silver nanostructures (as for Fig. 9C). Fig. 9 illustrates representative normalized (background subtracted) rotating disk voltammograms and ring currents recorded upon application of 1.21V during the reduction of oxygen for O$_2$- saturated 0.1 mol dm$^{-3}$ KOH at 1600 rpm rotation rate and 10 mVs$^{-1}$. Dotted lines stand for responses of the respective Ag-free graphene supports. Judging from the shape of the dotted disk and ring current-potential curves for all three Ag-free carbon supports (as for Fig. 9A, Fig. 9B and Fig. 9C) it can be concluded that two stages of limiting current can be obtained, indicating electrocatalytic reduction of oxygen proceeding via two-step processes, involving both two and four electron mechanisms with production of HO$_2^-$ intermediate. Judging from the shape of RDE responses recorded at the disk electrode covered with films containing silver-based catalysts it can be rationalized that only one stage of limiting current can be observed during the oxygen reduction process, suggesting the electrocatalytic reduction of oxygen proceeds is dominated by one-step processes involving four electron mechanisms. Moreover the performance on silver containing catalyst indicate that oxygen electroreduction process is mainly contributed by Ag while the carbon supports play the role of conducting, hydrophilic support improving the mobility of H$_2$O and OH$^-$ within the catalytic interface. Another important feature is that Gr/SiO$_2$ supported silver nanoparticles produces the lowest currents attributed to the production of hydrogen peroxide (lower than 6μAcm$^{-2}$) with the most positive onset potential (E$^o$) of the oxygen electroreduction process (as for Fig. 9A) when compared to the carboxylated-graphene supported silver nanoparticles (as for Fig. 9B) and reduced graphene oxide supported silver nanostructures (as for Fig. 9C). These experimental results clearly reveal that the silver nanoparticles deposited within the $SiO_2$-doted reduced-graphene-oxide (Gr/$SiO_2$) possesses pronounced catalytic activity for oxygen electroreduction process.

To get insight into the mechanism of oxygen reduction by monitoring the formation of intermediate peroxide species during ORR process the rotating ring-disk electrode (RRDE) voltammetric experiments have been performed. The percent amount of $HO_2^-$ (% $HO_2^-$) formed during reduction of oxygen under the conditions of RRDE voltammetric experiments of Fig. 9 have been calculated using the equation given below [43, 44]:

$$HO_2^- \% = \frac{2I_r/N}{I_d + I_r/N} \cdot 100\% \qquad (1)$$

where $I_r$ and $I_d$ are the ring and disk currents, respectively, and N is the collection efficiency. As shown in Fig.10, the $HO_2^-$ yield during ORR was 0.15%-1.96% at AgNPs/Gr/$SiO_2$, 0.30%-5.06% at carboxylated-graphene supported AgNPs and 0.44%-5.91% at AgNPs/Gr in a potential range from 0.05 to 0.64 vs RHE. The results clearly show that the production of $HO_2^-$ is the lowest for system utilizing silver nanoparticles deposited onto $SiO_2$-doted reduced-graphene-oxide (Gr/$SiO_2$), indicating that the ORR proceeds almost entirely through the 4e- pathway. Explanation of these phenomena should take into account the fact that structures of Ag NPs-supported Gr/$SiO_2$ are mainly surrounded with (111) facets (see XRD examination as for Fig. 5) that favor one-step "direct" four-electron reduction of oxygen directly to water. According to Wang [16] different ORR catalytic activity can be interpreted by the adsorption competition between oxygen and hydroxyl on different silver facets. The adsorption process of $O_2$ on the Ag (111) involve the bridge site of two neighbor silver atoms reaching the configuration similar to that on Pt (111), while during the adsorption of $O_2$ on the Ag (100), $O_2$ is settled on the hollow site confined by the under Ag atoms. The adsorption on Ag (100) surface is more effective than that on (111) surface. The increase of the HO* on surface limit the adsorption of $O_2$ and HOO*, due to decrease of the active sites on the surface caused by the coverage of HO. Therefore, the ratio of OH* coverage on Ag (100) is higher than that on Ag (111) due to the easy adsorption of OH* on Ag (100). The superfluous HO* formation on surface will block the formation of HOO* and be possible to lead to the desorption of $H_2O_{2ad}$. Thus, the possibility of two-electrons pathway is increased for Ag (100) surface.

The corresponding electron transfer number exchanged per $O_2$ molecule (n) was calculated as a function of the potential according to the RRDE voltammetric data from (Fig. 9 ), according to the equation given below [43,45,46]:

$$n = \frac{4I_d}{I_d + I_r/N} \qquad (2)$$

Remarkably, the oxygen electroreduction on $AgNPs/Gr/SiO_2$ yields the highest number of exchanged electrons (n) ranging from 3.96 to 3.998 when compared to the silver nanoparticles immobilized within carboxylated-graphene support (3.90-3.994) and reduced-graphene-oxide (Gr) supported silver nanoparticles (3.88-3.991) revealing the four-electron pathway is the main reaction process with the $H_2O$ as the product.

To further reveal the reaction kinetic of ORR the layer of $SiO_2$-rGO supported silver nanoparticles also was studied with rotating disk voltammetry. Fig. 12A shows a series of rotating disk voltammograms (RDVs) of ORR recorded on silver nanoparticles deposited within different carbon supports: carboxylated-graphene, $SiO_2$-doted reduced- at different rotation rates (from 225 to 3600 rpm) and potential sweep rate 10 mV s$^{-1}$ in an oxygen-saturated 0.1 M NaOH solution. The limiting currents of the ORR increase prominently with rotation rate, due to the enhanced oxygen diffusion to and reduction at the electrode surface but the onset potential of oxygen electroreduction is kept almost constant. The oxygen reduction is so fast that a limiting plateau is achieved proving that Ag (111) exhibit defect-rich surface coverage. The limiting currents at different rotation speeds were used to construct the Levich plot for different catalysts as shown for inset of Fig.12B, which is derived according to Koutecky-Levich equation [47-49]:

$$\frac{1}{i_{\lim}} = \frac{1}{i_k} + \frac{1}{i_L} = \frac{1}{C_{film} \cdot C_{O_2} \cdot k \cdot A \cdot n \cdot F} + \frac{1}{i_L} \qquad (3)$$

where $j_{lim}$ is the measured current density, $j_k$ and $j_L$ are the kinetic and convective- diffusion-limited current densities, respectively, n is the number of electrons transferred per $O_2$ molecule, k is the rate constant for $O_2$ reduction (in homogeneous units), F is Faraday constant (96485 C mol$^{-1}$), $C_{film}$ is the surface concentration of the catalytic sites, $C_{O2}$ is the bulk (solution) concentration of oxygen. For the Ag/m-MWCNT, the Koutecky-Levich plot ($j^{-1}$ vs. $w^{-1/2}$) exhibited a good linearity with an almost constant slope at potentials ranging from 0.05 V to 0.97 V (Fig. 12A).

It is necessary to consider the impact of any potential poisons or impurities on the catalytic activity of practical cathodic catalyst used in alkaline fuel cells (AFCs). Firstly, for alkaline electrolyte membrane fuel cells, the cathodic catalyst is expected be tolerant to the organic functional groups derived from anion exchange electrolyte membranes. Here, methanol and ethanol, as the most often studied fuels for direct alcohol fuel cells, are used for test. To comment on the observed phenomena, we have performed additional diagnostic voltammetric experiments, illustrating oxidations of organic fuels (methanol, ethanol, each at the 0.5 moldm$^{-3}$ level recorded in nitrogen-saturated 0.1 moldm$^{-3}$ KOH using a glassy carbon disk electrode covered with Gr/SiO$_2$ supported silver nanoparticles (Fig.13). Dash lines stand for responses of the respective alcohol-free solutions. It is apparent from Fig. 13A and 13B that no activity of AgNPs/Gr/SiO$_2$ toward the oxidation of methanol and ethanol is observed. AgNPs/Gr/SiO$_2$ therefore, proving high resistance of the catalytic systems towards potential contamination during oxygen reduction in the presence methanol or ethanol.

## 4. Conclusions

This study clearly demonstrates that the unique AgNPs/Gr/SiO$_2$ catalyst with a defect-rich surface was synthesized using Polyoxometallates (SiW$_{12}$O$_{40}^{4-}$) capping ligands assisted route and displays excellent catalytic performance for ORR in alkaline medium. The new synthetic method demonstrated in this study provides an efficient route for the generation of ultrafine and highly dense Ag nanoparticles that are homogeneously dispersed on different type graphene sheets: carboxylated-graphene, SiO$_2$-doted reduced- graphene-oxide (Gr/SiO$_2$) and reduced-graphene-oxide (Gr) and are sized about 5 nm. The XRD pattern of the obtained catalytic systems synthesized AgNPs are crystalline in nature. The total content of silver on different graphene supports was determined by the calibration curve ICP-MS and for all type of catalysts was on the level of 28% ± 4% (2*SD). The SiO$_2$-doted reduced- graphene-oxide supported silver nanoparticles exhibited pronounced electrocatalytic performance towards ORR with oxygen reduction peak at 0.81 V vs RHE when compared to carboxylated-graphene supporting silver nanoparticles and chemically reduced graphene oxide supporting Ag NPs with oxygen reduction peak at 0.76V. Moreover the nanocomposite based on the SiO$_2$-doted reduced- graphene-oxide supported silver nanoparticles produced the lower amount of H$_2$O$_2$ (0.2%-2%) under RRDe conditions and exhibited excellent methanol and ethanol tolerance.

**Acknowledgments**

We acknowledge the European Commission through the Graphene Flagship - Core 1 project [Grant number GA- 696656] and Maestro Project [2012/04/A/ST4/00287 (National Science Center, Poland)].

# Figure captions

**Figure 1.** Physicochemical characterization of samples: SiO$_2$-doted reduced- graphene-oxide (Gr/SiO$_2$) (red line), reduced-graphene-oxide (Gr) (black line): (a) HR-TGA profiles; (b) WAXD patterns; (c) confocal µ-Raman spectra.

**Figure 2.** Morphology of Gr (a); morphology of Gr/SiO$_2$ (b); HR-TEM studies of Gr/SiO$_2$: highly defective Gr nanoplatelets (c) and Gr covering the SiO$_2$ NPs (d).

**Figure 3.** Visible absorbance spectra of the following solutions (suspensions): solution of H$_4$SiW$_{12}$O$_{40}$ with SiO$_2$-doted reduced- graphene-oxide (a), solution of partially reduced H$_4$SiW$_{12}$O$_{40}$ (heteropolyblue) and SiO$_2$-doted reduced- graphene-oxide (b) and silver nanoparticles supported onto H$_4$SiW$_{12}$O$_{40}$-modified SiO$_2$-doted reduced- graphene-oxide (c).

**Figure 4.** Transmission electron microscopic images of (A) SiO$_2$-doted reduced- graphene-oxide (Gr-SiO$_2$) supported silver nanoparticles Ag (Gr/SiO$_2$) (30%w), (B) carboxylated-graphene-immobilized Ag nanoparticles (30%w) and reduced-graphene-oxide (Gr) supported Ag nanoparticle (30%w). Histograms A', B', and C' display distribution of sizes of the respective silver nanostructures.

**Figure 5.** XRD patterns of the representative samples of: (A) Gr/SiO$_2$ supported silver nanoparticles, (B) carboxylated-graphene supported silver nanoparticles and (C) reduced graphene oxide (Gr) supported silver nanostructures.

**Figure 6.** Cyclic voltammetric responses of (A) Gr/SiO$_2$ supported silver nanoparticles, (B) carboxylated-graphene supported silver nanoparticles and (C) reduced graphene oxide (Gr)

supported silver nanostructures recorded at 50 mVs$^{-1}$ in the de-oxygenated 0.1 moldm$^{-3}$ KOH. Dotted lines stand for responses of the respective Ag-free graphene-based supports.

**Figure 7.** Visible absorbance spectra of: (a) silver nanoparticles supported onto H$_4$SiW$_{12}$O$_{40}$-modified reduced- graphene-oxide and (b) reduced-graphene-oxide (Gr) supported silver nanoparticles, which were subsequently cleaned from their inorganic capping agents (PW$_{12}$O$_{40}$$^{4-}$) by converting to alkaline medium (0.1 mol dm$^{-3}$ KOH).

**Figure 8**. Background-subtracted linear scan voltammetric responses recorded at 10 mVs$^{-1}$ for the reduction of oxygen at the following catalytic layers: (A) Gr/SiO$_2$ supported silver nanoparticles, (B) carboxylated-graphene supported silver nanoparticles and (C) reduced graphene oxide (Gr) supported silver nanostructures. Electrolyte: oxygen-saturated 0.1 mol dm$^{-3}$ KOH. Silver loading: 30 mg cm$^{-2}$.

**Figure 9.** Normalized (background subtracted) rotating ring-disk voltammograms for oxygen reduction at (A) Gr/SiO$_2$ supported silver nanoparticles, (B) carboxylated-graphene supported silver nanoparticles and (C) reduced graphene oxide (Gr) supported silver nanostructures. Dotted lines stand for responses of the respective Ag-free graphene supports. Electrolyte: oxygen-saturated 0.1 moldm$^{-3}$ KOH. Scan rate: 10 mVs$^{-1}$ . Rotation rate: 1600 rpm. Ring currents are recorded upon application of 1.21V. Silver loading: 30 mgcm$^{-2}$ .

**Figure 10.** Percent fraction of hydrogen peroxide (% HO$_2$$^-$) produced during electroreduction of oxygen (and detected at ring at 1.21V) under conditions of the RRDE voltammetric experiments as for Fig. 9.

**Figure 11.** Numbers of transferred electrons (n) per oxygen molecule during electroreduction of oxygen under conditions of the RRDE voltammetric experiments as for Fig. 9.

**Figure 12.** (A) RDE polarization curves for ORR recorded in 0.1 M KOH at various rotation rates (225, 400, 900, 1699, 2500, 3600 rpm) and potential sweep rate 10 mV s$^{-1}$. (B)

Koutecky-Levich and Levich (Inset) plots obtained at (a) 0.6V and (B) 0.4V for Gr/SiO$_2$ supported silver nanoparticles catalyst using the data from RDE measurements of Fig. 12 A.

**Figure 13.** Cyclic voltammograms recorded for (A) C$_2$H$_5$OH and (B) CH$_3$OH oxidation (0.5 mol dm$^{-3}$ solutions) on catalytic film consisted of Gr/SiO$_2$ supported silver nanoparticles. Electrolyte: nitrogen-saturated 0.1 mol dm$^{-3}$ KOH. Scan rate: 10 mVs$^{-1}$. Dash lines stand for responses of the respective alcohol-free solutions.

Fig. 1

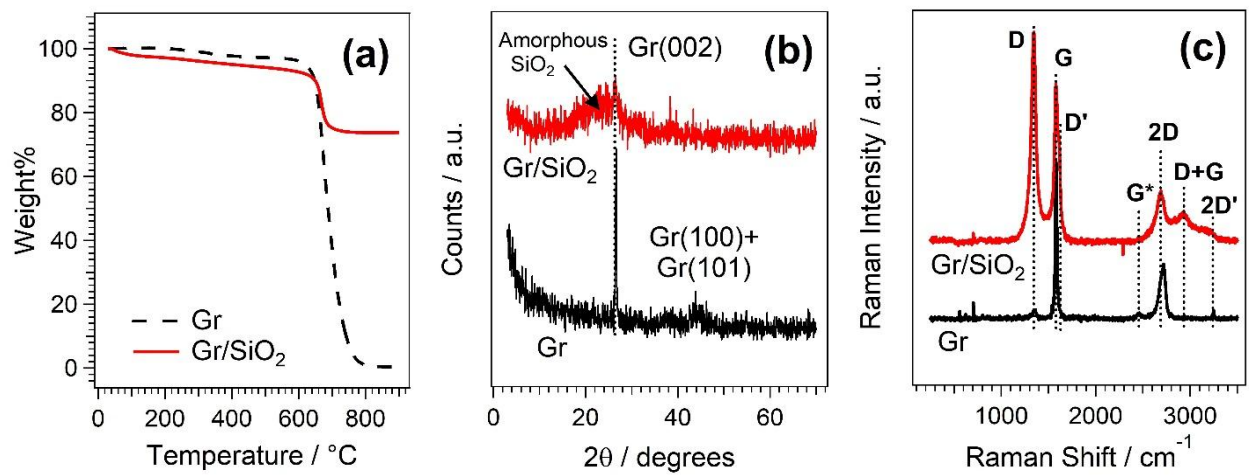

Fig.2

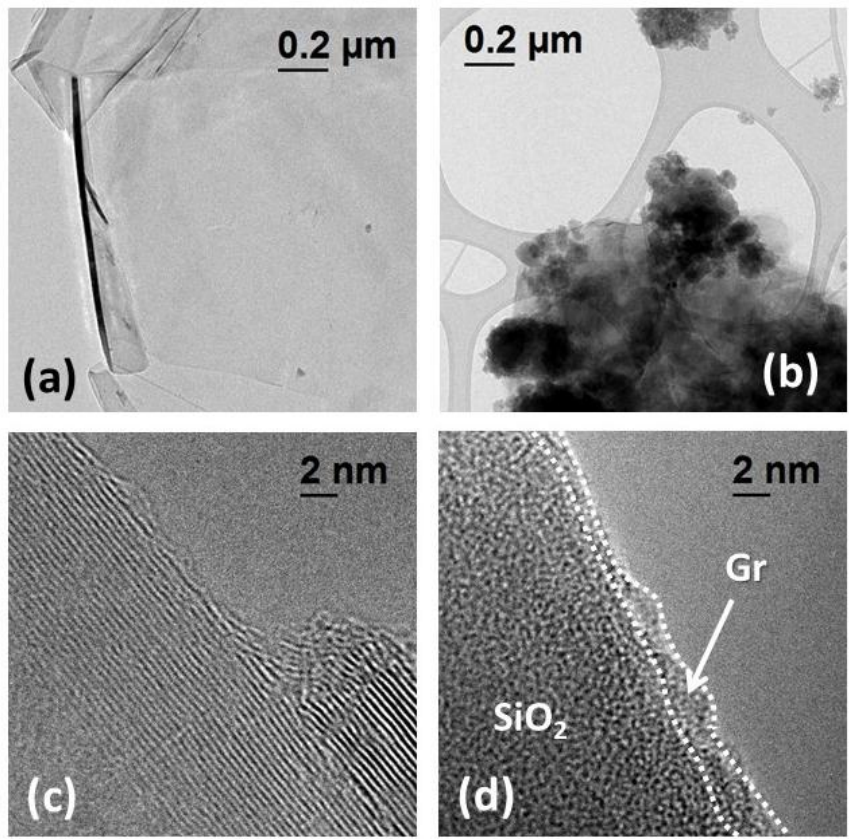

Fig.3

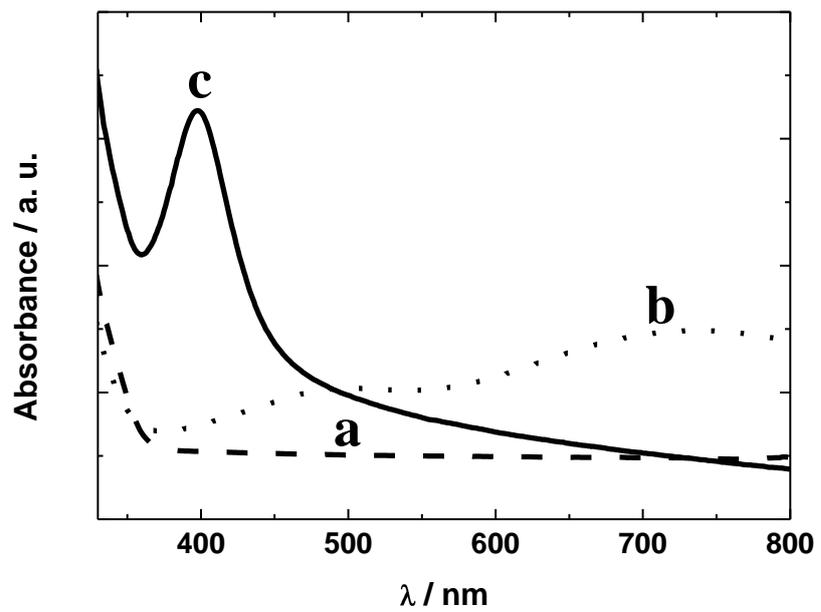

Fig.4

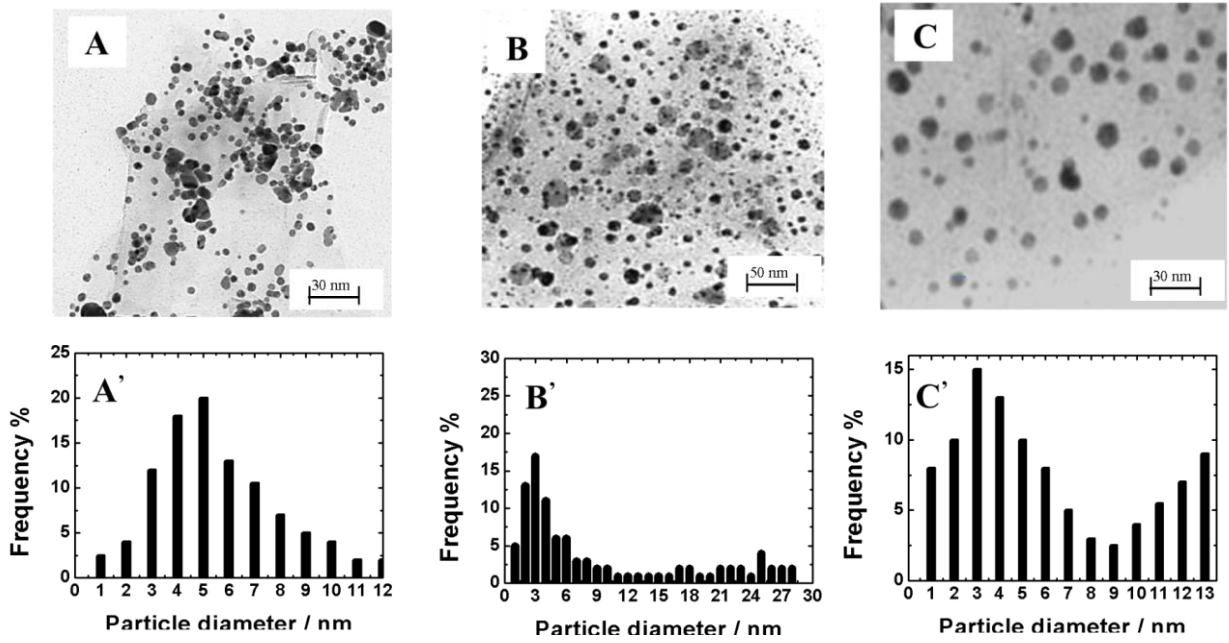

Fig.5

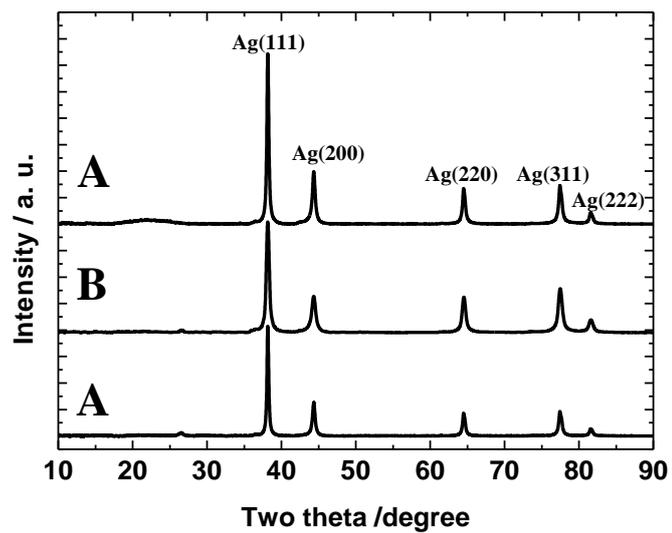

Fig.6

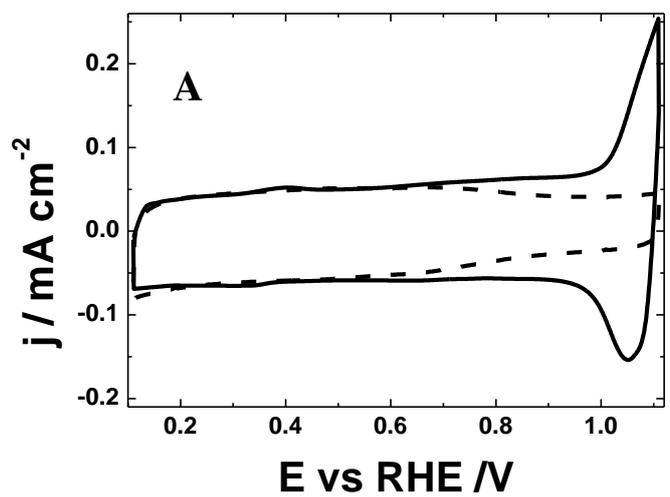

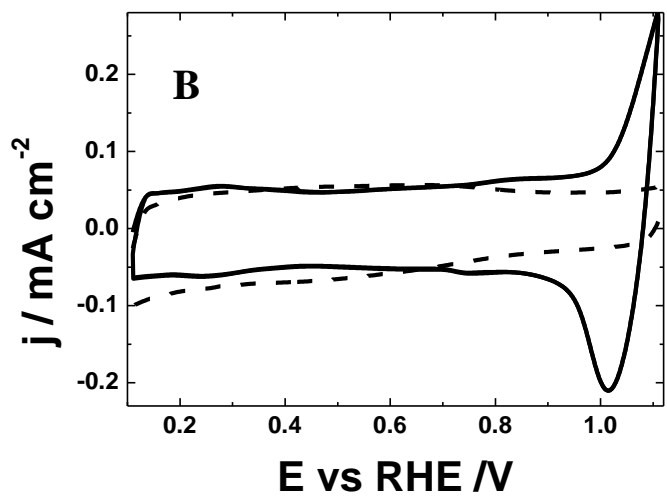

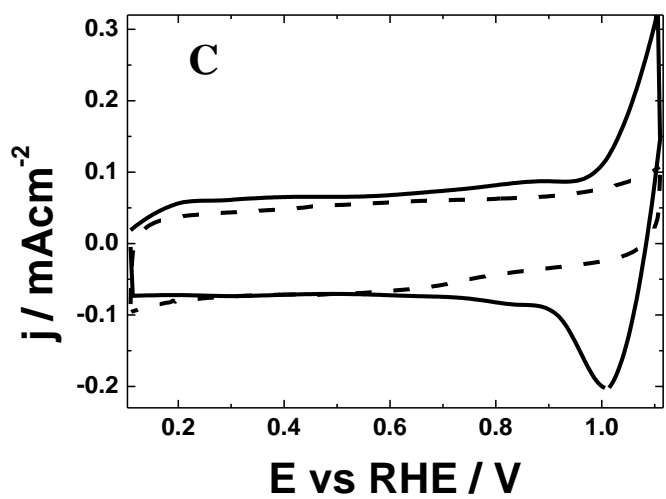

Fig.7

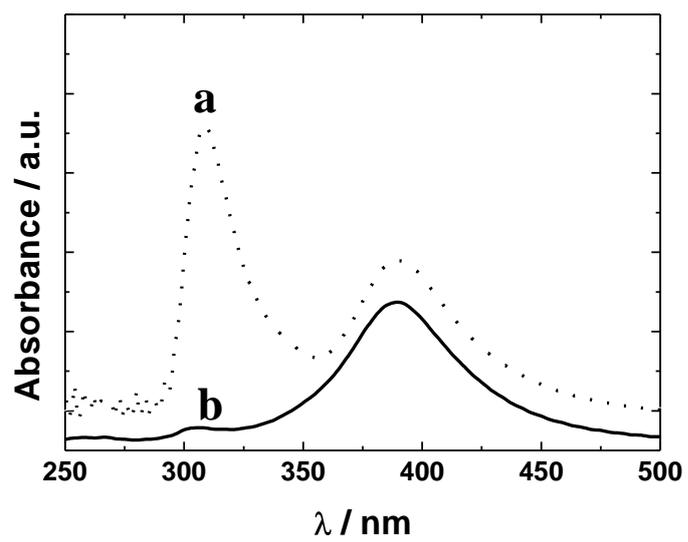

Fig.8

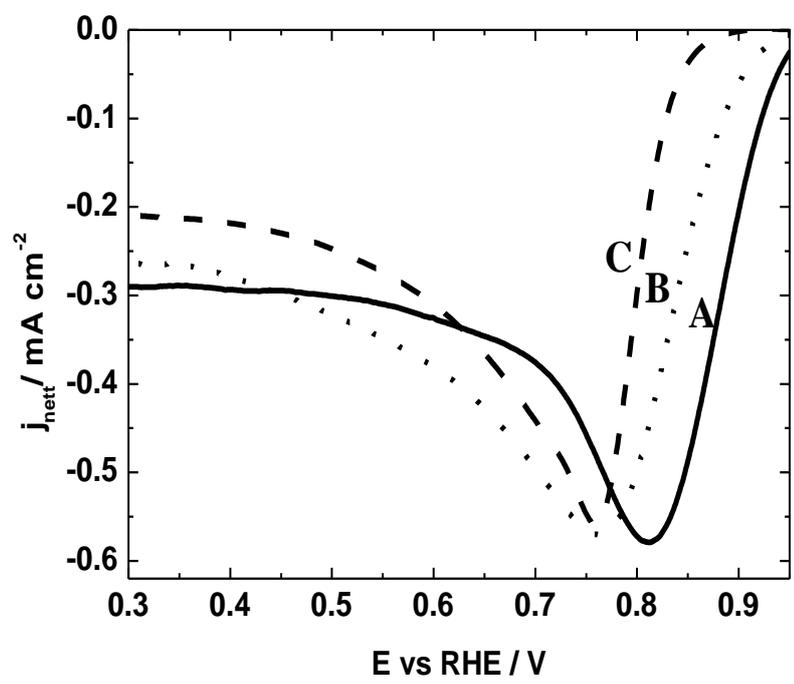

Fig.9

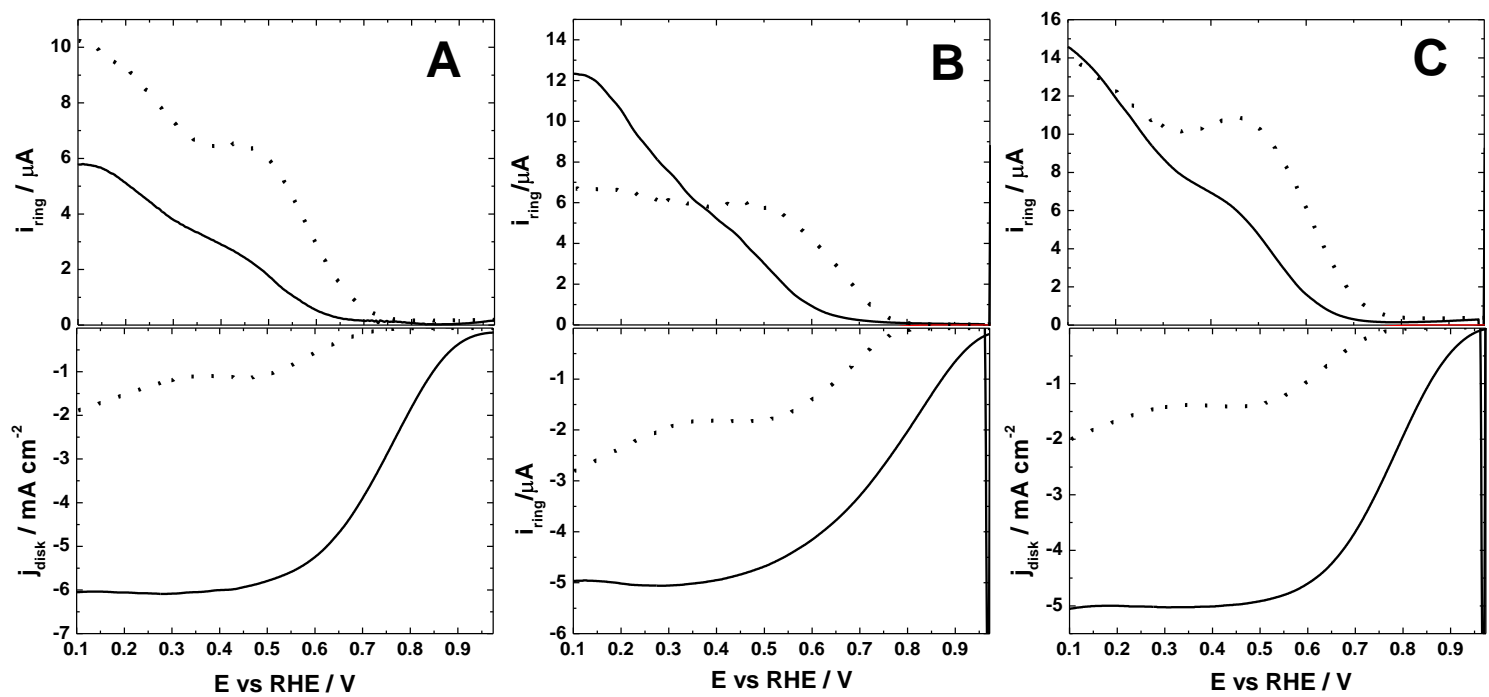

Fig. 10

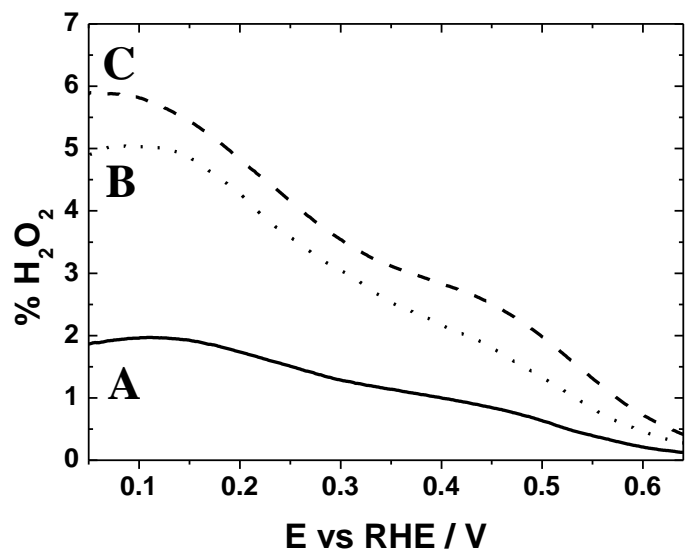

Fig. 11

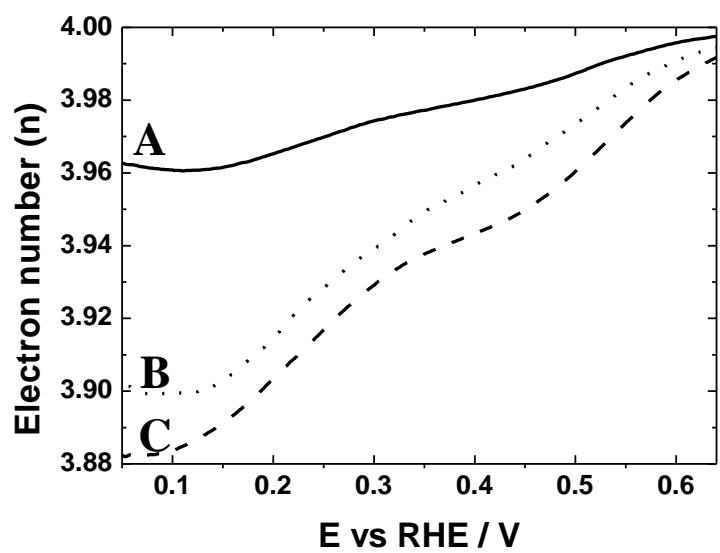

Fig. 12

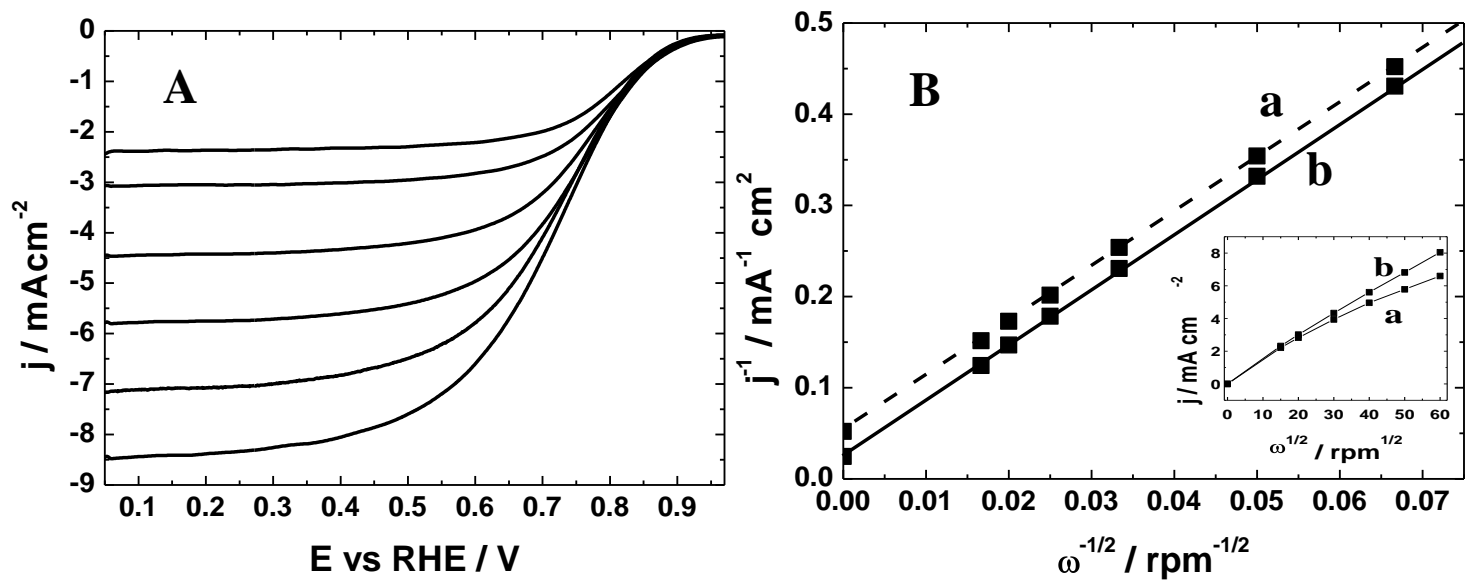

Fig. 13

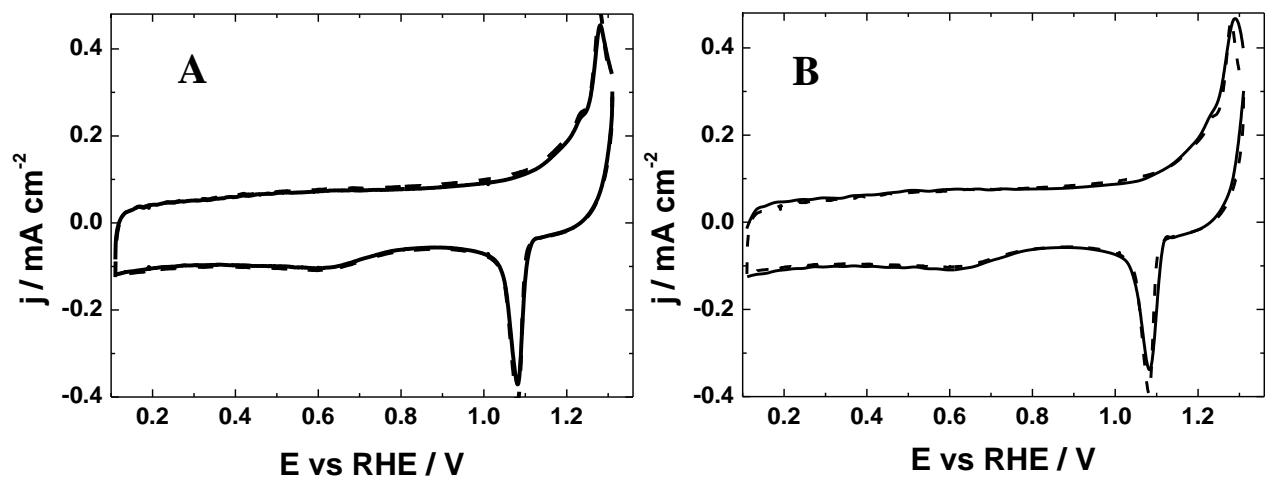

**Cartoon 1**

Schematic diagram illustrating the preparation of graphene-supported silver nanoparticles capped with Keggin silicotungstic heteropolyacid.

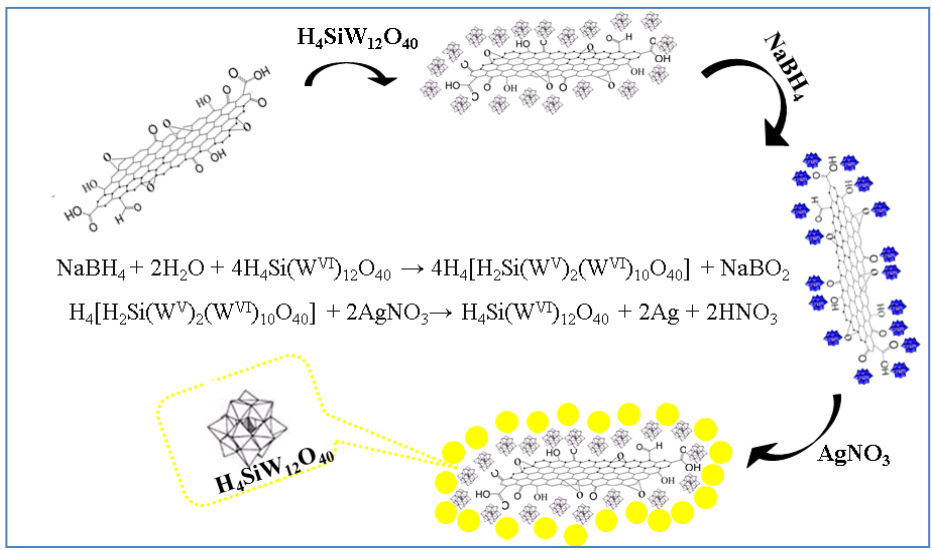

$NaBH_4 + 2H_2O + 4H_4Si(W^{VI})_{12}O_{40} \rightarrow 4H_4[H_2Si(W^V)_2(W^{VI})_{10}O_{40}] + NaBO_2$

$H_4[H_2Si(W^V)_2(W^{VI})_{10}O_{40}] + 2AgNO_3 \rightarrow H_4Si(W^{VI})_{12}O_{40} + 2Ag + 2HNO_3$